# AI for Earth: Rainforest Conservation by Acoustic Surveillance

Yuan Liu
Huawei Cloud
liuyuan45@huawei.com

Zhongwei Cheng
Futurewei
zhongwei.cheng@futurewei.com

Jie Liu
Huawei Cloud
roger.liujie@huawei.com

Bourhan Yassin
Rainforest Connection
bourhan@rfcx.org

Zhe Nan
Futurewei
zhe.nan@futurewei.com

Jiebo Luo
Futurewei
jiebo.luo@futurewei.com

## ABSTRACT

Saving rainforests is a key to halting adverse climate changes. In this paper, we introduce an innovative solution built on acoustic surveillance and machine learning technologies to help rainforest conservation. In particular, We propose new convolutional neural network (CNN) models for environmental sound classification and achieved promising preliminary results on two datasets, including a public audio dataset and our real rainforest sound dataset. The proposed audio classification models can be easily extended in an automated machine learning paradigm and integrated in cloud-based services for real world deployment.

## KEYWORDS

audio classification, acoustic surveillance, neural networks

## 1 INTRODUCTION

Rainforests are the Earth's oldest living ecosystems, which over half of all biotic species are indigenous to. Deforestation contributes to nearly 1/5 of all global carbon emission. Therefore, protecting reforests is a key to fighting against climate changes and preserving bio-diversity. Tremendous efforts have been made to help save rainforests with modern technologies. With the increasing successful applications of machine learning and artificial intelligence in various industries and society in general, such advanced technologies are also attracting the attention of people who are dedicated to protecting our planet.

We focus on the task of rainforest conservation and attempt to bring AI into the picture to significantly improve the efficiency and effectiveness of the protection on the ground. Despite the recent successes of such technologies in many domains, applying them to real-world conservation projects remains challenging. Deploying AI-powered systems in rural areas faces a wide variety of critical restrictions, such as very limited power supply, poor connectivity, and harsh conditions. We need to utilize a practical yet effective modality to sense the environment and provide informative data to support decision making. Compared with image data which is limited by field of view and area covered with dense understory such as the rainforest, audio data can be a good fit in the sense of ease and robustness of data acquisition, low data volume and high information density. Several early attempts [9] of acoustic surveillance for rainforests have reported the promising value of audio information.

Adapting audio recognition techniques in rainforest conservation tasks such as detecting illegal deforestation, animal poaching, as well as bio-acoustic monitoring are nontrivial, although urban sound classification models have outperformed humans. The main challenges come from the notable domain gap between the sounds of urban and natural environments, and the shortage of extremely expertise-dependent labeled data. Fig.1 illustrates the Log-Mel spectrums of chainsaw sounds sampled from different sources. It shows the significant characteristic differences in the same kind of sounds in different scenarios caused by the variations in the background sounds and distances from the sensing devices.

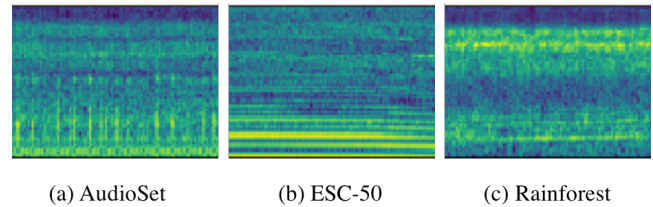

(a) AudioSet  (b) ESC-50  (c) Rainforest

Figure 1: Log-Mel spectrograms of chainsaw sounds from various sources.

We introduce our preliminary work on improving audio analysis towards rainforest conservation. In addition, we briefly outline our vision on building cloud AI powered conservation systems to bridge the domain experts and AI specialists while making it easy to enable technology utilization for good.

## 2 AUDIO RECOGNITION

Convolutional neural network (CNN) based models [2, 5, 8] have been proposed for sound event detection (SED). While they have achieved very promising results on public datasets for research, we cannot directly employ these existing models or labeled resources since there is significant domain gap between them and our targeted rainforest tasks. Computational efficiency is also a concern.

To address the above concerns, we propose two new modified CNN models to better balance model capacity and capability, and also to improve model performance using transfer learning techniques to leverage large-scale weakly labeled audio datasets.

### 2.1 Augmented VGGish model

The VGGish model [2] broadly used in audio recognition is modified from VGG16 with layer pruning to reduce the number of parameters. However, it still holds 72.1M parameters and its model (in)efficiency restricts its application on resource-limited devices like IoT devices. Moreover, VGGish only supports single sized input features. In order to accommodate various input sizes to achieve flexibility in transfer learning and better performance, we propose an augmented VGGish network. The Aug-VGGish model includes only 4.7M parameters and is able to achieve better recognition performance. The modifications we made are as follows:

1. Batch Normalization[3] is introduced following each convolutional layer. Batch Normalization allows much higher learning rates and is less sensitive to initialization.

2. A global pooling layer replaces a flattened layer. Adding the global pooling layer not only helps filter the features but also makes the network adapt to different sizes of input spectrograms.

3. The 4096-unit FC layers are reduced to 256-unit FC layers.

4. The final 128-unit dense layer is removed.

## 2.2 Fully convolutional network

The Aug-VGGish model produces good performance in our experiments, while global pooling still tends to lose potential informative signals. Given that fully convolutional networks (FCNs) [4] have shown leading performance in vision tasks, we propose to employ FCN in audio recognition. Our proposed FCN-VGGish model contains 8 convolutional layers with 18.7M parameters in total. The FCN-VGGish network has the advantages of Aug-VGGish models in addition to the enhanced capability brought by more convolutions.

## 3 PRELIMINARY RESULTS

We conduct preliminary experiments to validate our proposed models. We test our models on audio classification tasks against two datasets, one of which is a public benchmark audio dataset and the other is real-world audio samples collected in rainforests.

Our models are first trained on weakly labeled AudioSet [1], and then transferred to the target tasks with parameter fine-tuning. Performance results of difference models are reported with an identical training setup unless stated otherwise.

### 3.1 Results on the ESC-50 dataset

ESC-50 [6] is a balanced public dataset that contains 2000 audio recordings of 50 classes, 40 clips per class and 5 seconds per clip. Results based on 5-fold cross-validation are listed in Table1 to evaluate the model performance. Taking human capability of 81.3% accuracy on this task as a reference, the current state-of-the-art [7] listed on the dataset webpage is 86.5%. Our proposed models obtain clearly better performance with 87.5% accuracy for Aug-VGGish and 90.1% accuracy for FCN-VGGish. Comparing to the vanilla VGGish model, our modified new models achieve at least 6.2% improvement, which validates the effectiveness of our modifications.

Table 1: Audio classification performance on ESC-50

| Model | Mean Accuracy | F1 score |
| --- | --- | --- |
| Human Accuracy [6] | 81.3% | N/A |
| FBEs+ConvRBM-BANK[7] | 86.5% | N/A |
| VGGish[2] | 81.3% | 0.806 |
| *Aug-VGGish* | **87.5%** | **0.870** |
| *FCN-VGGish* | **90.1%** | **0.898** |

### 3.2 Results on real-world sounds in rainforests

The rainforest environmental audio data was collected on-site and annotated by our NGO partner Rainforest Connection. Note that used Huawei smartphones are deployed as the sensors due to their long battery life and robustness in the harshly hot and wet conditions of the rainforest. This dataset currently includes 22000 audio recordings with annotations of chainsaw or not per one second clip, and is very unbalanced between the 2 classes. The target chainsaw sounds are notably different from those in ESC-50 as shown in Fig.1.

We test the VGGish model and our proposed variations on this dataset with the same transfer learning strategy. The performance comparison of classifying chainsaw sounds in rainforests is shown by the precision-recall curves in Fig2. Note that there are no extreme samples like insect buzzing in the current data collection, therefore the overall performance seems satisfactory. We can see that Aug-VGGish outperforms vanilla VGGish and FCN-VGGish is the best.

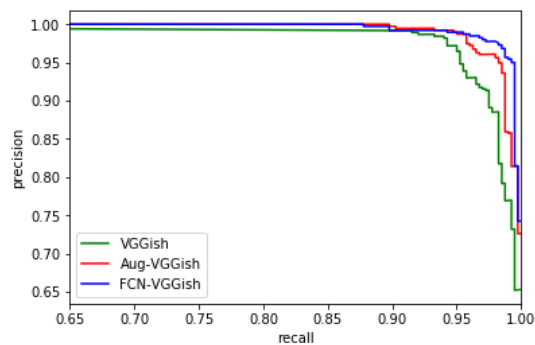

Figure 2: Chainsaw sound classification in rainforests.

## 4 FUTURE WORK

We will explore transfer learning and few-shot learning techniques to improve audio recognition in natural environments, including further rainforest conservation tasks such as spider monkey habitat modeling and monitoring. With our NGO partner, we are making cloud-based AI solutions for rainforest conservation a reality.